\documentclass[useAMS,usenatbib,a4paper, fleqn]{mnras}
\usepackage{ae,aecompl}
\title[MHD models of colliding clouds]
{The properties of clusters, and the orientation of magnetic fields relative to filaments, in magnetohydrodynamic simulations of colliding clouds
}
\author[Dobbs]
{C. L. Dobbs\thanks{E-mail:
C.L.Dobbs@exeter.c.uk}$^{1}$, J. Wurster$^{2}$ \\
$^1$ School of Physics and Astronomy, University of Exeter, Stocker Road, Exeter, EX4 4QL, UK \\
$^2$ Scottish Universities Physics Alliance (SUPA), School of Physics and Astronomy, University of St. Andrews, North Haugh, \\
St Andrews, Fife KY16 9SS, UK \\}

\usepackage{amssymb}
\usepackage{amsmath}
\usepackage[pdftex]{graphicx}
\usepackage{epsfig}
\usepackage{multirow} 
\usepackage{bm}           
\def\aap{{ A\&A}}

\def\apjl{{ApJL}}

\def\apjs{{ApJS}}

\begin{document}
\label{firstpage}
\date{\today}

\pagerange{\pageref{firstpage}--\pageref{lastpage}} \pubyear{2012}

\maketitle

\begin{abstract}
We have performed Smoothed Particle Magneto-Hydrodynamics (SPMHD) calculations of colliding clouds to investigate the formation of massive stellar clusters, adopting a timestep criterion to prevent large divergence errors. We find that magnetic fields do not impede the formation of young massive clusters (YMCs), and the development  of high star formation rates, although we do see a strong dependence of our results on the direction of the magnetic field. If the field is initially perpendicular to the collision, and sufficiently strong, we find that star formation is delayed, and the morphology of the resulting clusters is significantly altered. We relate this to the large amplification of the field with this initial orientation. We also see that filaments formed with this configuration are less dense. When the field is parallel to the collision, there is much less amplification of the field, dense filaments form, and the formation of clusters is similar to the purely hydrodynamical case. Our simulations reproduce the observed tendency for magnetic fields to be aligned perpendicularly to dense filaments, and parallel to low density filaments. Overall our results are in broad agreement with past work in this area using grid codes.
\end{abstract} 


\section{Introduction}
Recent works have shown that young massive clouds (YMCs) can form through the collision of molecular clouds \citep{Dobbs2020,Liow2020}. \citet{Dobbs2020} showed that YMCs are able to form on timescales of 1-2 Myr, in line with observed age spreads \citep{Longmore2014}. Observationally, there is evidence of cloud cloud collisions in our Galaxy from red and blue shifted CO velocities in molecular clouds along the line of sight, in some cases at the sites of massive young clusters \citep{Furukawa2009,Fukui2014,Fukui2017,Kuwahara2020}. \citet{Dobbs2015} also found in galaxy scale simulations that such collisions of massive clouds, although infrequent, do occur. \citet{Liow2020} carried out a parameter study showing high density, low turbulence and high velocities promote YMC formation. They also determined the properties of clusters which formed, showing that for the cloud masses used ($10^4-10^5$ M$_{\odot}$) the properties are comparable to lower mass YMCs in our Galaxy \citep{PZ2010}, though high velocities led to more elongated clouds and larger cloud radii, at least in the earliest stages of evolution. These previous simulations however are all purely hydrodynamical. Whether such clusters still form when magnetic fields are present, and still have the same properties, is an open question.

A number of studies have examined the effects of magnetic fields in simulations of colliding flows of interstellar gas \citep{Heitsch2009,Inoue2009,Kortgen2015,Wu2017,Wu2020,Klassen2017,Fogerty2016,Fogerty2017,ZA2018,Seifried2020}. \citet{Heitsch2009} carry out simulations investigating molecular cloud formation, and show that there is a strong dependence on the initial field direction, with the collision only inducive to producing molecular clouds when the field is parallel to the direction of flow. More recent works have included self gravity and shown the impact of magnetic fields on core and star formation. \citet{Kortgen2015} find that magentic fields delay core and star formation, although \citet{ZA2018} find that star formation occurs earlier with strong magnetic fields, due to suprresion of the non-linear thin shell instability. \citet{Sakre2020} model core formation and suggest that stronger fields provide support to allow the formation of more massive cores (see also \citealt{Inoue2018}). \citet{Sakre2020} also investigate field direction and find that starting with an initial field parallel to the collision produces a more disordered field compared to when the initial field is perpendicular. \citet{Wu2020} also find that less fragmentation occurs in models with stronger fields, and there are fewer stars. Although some studies now explicitly include sink particles 
\citep{Inoue2018,Fogerty2016,ZA2018,Fukui2020}, few investigate the role of magnetic fields on cluster formation.

Filaments are widespread both in the neutral ISM and star forming regions, and as such many of the above studies also investigate the relation of magnetic fields to filaments. Recent observations now reveal the alignment of magnetic fields with structures in the ISM. In particular, observations appear to show that the magnetic field is typically aligned parallel to filaments in HI  \citep{McClure-Griffiths2006,Clark2014,Planck2016a}, and low density molecuar gas \citep{Heyer2012,Heyer2020}. Whereas in higher density molecular gas, the field is more likely to be perpendicular to the filaments \citep{Alves2008,Heyer2012,Planck2016b}. \citet{Fissel2019} show examples of both parallel and perpendicular alignment within the Vela cloud, traced by $^{12}$CO and $^{13}$CO respectively. Simulations of turbulence \citep{Soler2013,Klassen2017,Xu2019} and shock compressed layers \citep{Chen2016} also show a tendency for the magnetic field to be aligned perpendicular to the field at high density, and parallel otherwise. The dependence of the field orientation may be simply a density criterion \citep{Soler2013,Chen2016}, or additionally related to the mass to flux ratio \citep{Seifried2020}. \citet{Inoue2016} find that the alignment of the magnetic field with the filament depends on the initial angle between the shock wave and the magnetic field. 

In this paper we perform simualtions of colliding clouds with magnetic fields, though our focus is on the formation of massive clusters rather than individual stars. In Section \ref{sec:method} we describe our method and initial conditions, and in particular the timestep constraint we apply to ensure the magnetic divergence remains low. We descibe the morphologies of the collisions, star formation rates, the relation of magnetic field to filaments and the properties of clusters formed in Section~\ref{sec:results}. In Section~\ref{sec:comparison} we compare to previous work, and in Section~\ref{sec:conclusions} we present our conclusions.

\section{Method}\label{sec:method}
\subsection{Details of Simulations}
We have performed these calculations using \textsc{Phantom} \citep{Price2018}, which is a publicly available Smoothed Particle Magneto-Hydrodynamics (SPMHD) code. Sink particles are included according to the method described in \citet{Bate1995}. Magnetic fields are evolved as the magnetic variable $B/\rho$.  Stability of the magnetic fields is ensured using the source term correction \citep{Borve2001} and the divergence is constrained using hyperbolic divergence cleaning \citep{Tricco2012,Tricco2016}.  
Unlike previous work, we apply a modified timestep criteria based upon the divergence cleaning method, which we describe in Section~\ref{sec:timestep}. This timestep constraint, which is in addition to the usual Courant and acceleration criteria \citep{Price2018}, ensures that the timesteps are small enough to prevent large increases in the divergence. For simplicity, and so that when we vary the velocity field or magnetic field everything else is unchanged, we employ an isothermal equation of state, adopting a temperature of 20 K. 

We set up the initial conditions for our simulations in a similar way to \citet{Dobbs2020} and \citet{Liow2020}, though with a few differences. We simulate two ellipsoidal colliding clouds, which are colliding head on along their minor axes. The clouds have dimensions of $80 \times 30 \times 30$ pc. Both clouds have masses of $1.5\times 10^5$ M$_{\odot}$. All the simulations we present use 6.1 million particles. Although not shown, we initially ran simulations with one tenth the number of particles, which show very similar results to those we display here. The initial setup of the clouds differs from our previous simulations in two main ways. The first is that the two clouds lie within a low density medium, which is one hundredth of the density of the clouds. This is the same approach as \citet{Wurster2019}, who modelled isolated clouds within a low density medium. The magnetic field permeates both the clouds and this surrounding medium, preventing the magnetic field becoming unstable at the edge of the cloud, and removing the need for more complex magnetic boundary conditions. For simplicity, the surrounding medium has periodic boundary conditions (satisfying the need for magnetic boundaries), where magnetohydrodynamic forces are periodic across the boundary but gravitational forces are not.  Including the low density medium ensures that any increase in the field does not occur at the edge of the simulation region, since the field does not evolve significantly in the low density region. The extent of the low density medium is $\pm 120$ pc in the $x$ dimension, and $\pm 46$ pc in the $y$ and $z$ dimensions. Secondly, following the setup of the initial conditions in \citet{Wurster2019}, the particles are initially allocated on a grid in both the clouds and the low density surrounding medium, rather than randomly.

\begin{table*}
\begin{tabular}{c|c|c|c|c}
 \hline 
Run & $\sigma$  & Virial & Magnetic field  & Direction of field  \\
  &  (km s$^{-1}$) & parameter ($\alpha$) & strength (G)  & relative to collision \\
   \hline
BWXLowturb & 2.5 & 0.4 & $2.5\times10^{-7}$ & parallel \\
BSXLowturb & 2.5 & 0.4 &  $2.5\times10^{-6}$ & parallel  \\
BWYLowturb & 2.5 & 0.4 &  $2.5\times10^{-7}$ & perpendicular \\
BSYLowturb & 2.5 & 0.4 &  $2.5\times10^{-6}$ & perpendicular\\
HydroLowturb & 2.5 & 0.4 & 0 & - \\
BWXHighturb & 5 & 1.7 &  $2.5\times10^{-7}$ & parallel  \\
BSXHighturb & 5 & 1.7 &  $2.5\times10^{-6}$ & parallel  \\
BWYHighturb & 5 & 1.7 &  $2.5\times10^{-7}$ & perpendicular \\
BSYHighturb & 5 & 1.7 &  $2.5\times10^{-6}$ & perpendicular \\
HydroHighturb & 5 & 1.7 & 0 & - \\
\hline
\end{tabular}
\caption{Table showing the initial configurations for the simulations performed in this paper. The virial parameter is calculated as the ratio of the kinetic energy to the gravitational potential energy.}
\label{tab:simulations}
\end{table*}

As for the previous simulations \citep{Liow2020}, we apply a turbulent velocity field to each cloud. The velocity field is set up to follow a Gaussian distribution, which produces a power spectrum consistent with $P(k)\propto k^{-4}$, Burger's turbulence. In our previous work, we showed that quite high velocities are required to form massive clusters over short timescales. 
Here we only consider one set of collision velocities, and set up each cloud with a velocity of 21.75 km s$^{-1}$, such that the total relative velocity between the clouds is 43.5 km s$^{-1}$. This velocity is chosen so that the collision has a significant effect on star formation, i.e. the collision enhances the star formation rate above that which would occur for isolated clouds \citep{Dobbs2020}, but is still consistent with the highest velocities observed for colliding streams in the Milky Way \citep{Motte2014,Fukui2015,Fukui2018}. In \citet{Liow2020} we show results with different cloud dimensions and collision velocities, but here we focus on varying the magnetic field strength and orientation. 

We apply two different field strengths, of $2.5\times10^{-6}$ and  $2.5\times10^{-7}$ G, and align these fields either parallel or perpendicular to the collision. 
We note that particularly our weaker field strength is unrealistically low compared to observations, but is intended for comparisons. Our higher field strengths are at the low end of observed field strengths (e.g. \citealt{Heiles2005,Crutcher2010}). We discuss in Section \ref{sec:conclusions} how we expect the trends we observe to extend to higher strengths. The Alfv\'{e}n velocity is $\sim$0.06 and 0.6 km s$^{-1}$ for the weak and strong fields respectively, so significantly lower than both the turbulent velocity field, and collision velocity.
We also carry out purely hydrodynamical simulations as well for comparison. We vary the velocity dispersion of the turbulence, which produces clouds which are unbound and bound initially. We show the simulations presented in Table~\ref{tab:simulations}. As indicated in Table~\ref{tab:simulations} the main variables in the simulations are the level of turbulence, which changes the virial parameter, the magnetic field strength and the orientation of the magnetic field.

We insert sink particles once the density reaches a critical density of $10^{-18}$ g cm$^{-3}$ and the criteria in \citet{Bate1995} (e.g. converging flows) are fulfilled, using an accretion radius of 0.001 pc. With this resolution, each sink particle typically represents a small group of stars. Artificial viscosity is included with a switch for the $\alpha$ parameter \citep{Cullen2010}. As recommended for strong shocks \citep{Price2010}, we take $\beta=4$. The artificial resistivity is described in \citet{Price2018}.

\subsection{Divergence cleaning timestep contraint}\label{sec:timestep}
The magnetic field is evolved as
\begin{equation}
    \rho\frac{\text{d}}{\text{d}t}\left(\frac{\bm{B}}{\rho}\right) = \left(\bm{B}\cdot\bm{\nabla}\right)\bm v - \nabla\psi,
\end{equation}
where $\rho$ is the gas density, $\bm{B}$ is the magnetic field, $\bm{v}$ is the velocity and $\psi$ is a scalar field used for divergence cleaning.  We assume units of the magnetic field such that the Alfv{\'e}n speed is $v_\text{A} = \left|\bm{B}\right|/\sqrt{\rho}$ \citep{Price2004}.  As per \citet{Tricco2016}, the evolved cleaning parameter is $\psi/c_\text{h}$, where $c_\text{h}$ is the characteristic speed, referred to as the `wave cleaning speed.'  The evolution of the parameter is given by\footnote{We have modified this slightly from \citet{Tricco2016} to explicily include the overcleaning parameter $\sigma$, which has a slightly different definition here.  We explicitly note that this $\sigma$ is not the velocity dispersion.}
\begin{equation}
\label{eq:clean}
     \frac{\text{d}}{\text{d}t}\left(\frac{\psi}{c_\text{h}}\right) = -\sigma c_\text{h}\left(\bm{\nabla}\cdot\bm{B}\right) - \frac{\sigma c_\text{h}}{h} \left(\frac{\psi}{c_\text{h}}\right) - \frac{1}{2}\left(\frac{\psi}{c_\text{h}}\right)\left(\bm{\nabla}\cdot\bm{v}\right),
\end{equation}
where $c_\text{h} = \sqrt{v_\text{A}^2 + c_\text{s}^2}$ with $c_\text{s}$ being the sound speed and $h$ is a scale length (equal to the smoothing length in SPH).  To ensure that the cleaning is resolved, a new timestep criteria is introduced.  In SPH, the new timestep for particle $i$, given by
\begin{equation}
\label{eq:dtclean}
    \text{d}t_{\text{clean},i} = \frac{C_\text{cour}h_i}{2\sigma_{ij}c_{\text{h},i}},
\end{equation}
where $C_\text{cour} = 0.3$ is the tradional coefficient for the Courant condition.

\citet{Tricco2016} introduced the `overcleaning' parameter $\sigma_{ij} \equiv \sigma$ to control the cleaning. Optimally, $\sigma = 1$, however, larger values could be chosen to reduce divergence errors, albeit at the accompying cost of shorter timesteps (recall \eqref{eq:dtclean}).  The divergence error is monitored by the dimensionless value
\begin{equation}
    \epsilon_\text{divB} = \frac{h |\bm{\nabla} \cdot \bm{B}|}{|\bm{B}|},
\end{equation}
and they suggest increasing $\sigma$ if the mean value is $>10^{-2}$. 

In most simulations where the magnetic field is reasonably well-behaved \citep[e.g.][]{OrszagTang1979,RyuJones1995,Wurster2019}, $\left.\epsilon_\text{divB}\right|_\text{mean} < 10^{-2}$ is satisfied using the default value of $\sigma = 1$.  However, in \textbf{very dynamic} regions, such as the interface between colliding flows as presented here, this criteria is violated; away from the interface, however, this criteria is satisfied.  Therefore, $\sigma > 1$ is required for the stability of the magnetic field at the interface.  This requires a careful choice of $\sigma$ prior to starting the simulation such that it is large enough to properly clean the magnetic divergence, but low enough such that computational resources are not wasted.

To circumvent this and to prevent extra computational expense away from the interface, we dynamially calculate $\sigma_{ij}$ based upon a particle's local environment.  Specifically,
\begin{equation}
    \sigma_{ij} = \min\left[{\sigma_\text{max}, \max\left(\sigma,fh_i\left.\frac{\left|\bm{\nabla}\cdot\bm{B}\right|}{\left|\bm{B}\right|}\right|_i,fh_j\left.\frac{\left|\bm{\nabla}\cdot\bm{B}\right|}{\left|\bm{B}\right|}\right|_j\right)}\right],
\end{equation}
where $\sigma_\text{max}$ is a parameter defining the maximum permitted $\sigma_{ij}$, $f$ is a scalar, and this minimising operation occurs over all of $i$'s neighbours, particles $j$.
Note that we keep a constant value of $\sigma$ in the wave cleaning equation, \eqref{eq:clean}.  The advantage to this method is that we do not need to guess a value of $\sigma$ prior to starting the simulation, and $\sigma_{ij}$ will only increase as needed and where it is needed, which will save computational resources given our use of individual timestepping \citep{Bate1995}.
Emperical tests of colliding flows similar to those presented here suggest $f = 10$ and $\sigma_\text{max} = 512$.

Although $\sigma_{ij}$ is dynamically calculated, careful consideration must be made of $\sigma$ and $\sigma_\text{max}$.  In the Ryu \& Jones MHD shock tube tests \citep{RyuJones1995}, $\left.\epsilon_\text{divB}\right|_\text{max} < 10^{-2}$ meaning that the new algorithm has no impact and it is safe to use the default values.  However, in the Orszag–Tang vortex \citep{OrszagTang1979} with 128 particles in the $x$-direction, the mean value of $\epsilon_\text{divB}$ is $\lesssim 0.005$, while  $\left. \epsilon_\text{divB}\right|_\text{max} \sim \mathcal{O}(1)$.  In this test, setting $\sigma_\text{max} = 512$ has a trivial affect on the results (including the value of $\left.\epsilon_\text{divB}\right|_\text{max}$), except the simulation runs $\approx10$ times slower when performed with global timestepping due to the decreased d$t_\text{clean}$; in this case, it is optimal to set $\sigma_\text{max} = 1$, essentially turning off the dynamic calculation of $\sigma_{ij}$.  

Therefore, this new timestep criterion is required when modelling magnetic fields at \textbf{strong, chaotic shocks} (such as the colliding flows presented here) to permit a reliable evolution of the magnetic field.  However, it should be disabled when modelling well-behaved magnetic fields.

\section{Results}\label{sec:results}
\subsection{Morphology of shocked region, clusters and magnetic field}
\begin{figure*}
\centerline{\includegraphics[scale=0.4, bb=450 0 800 650]{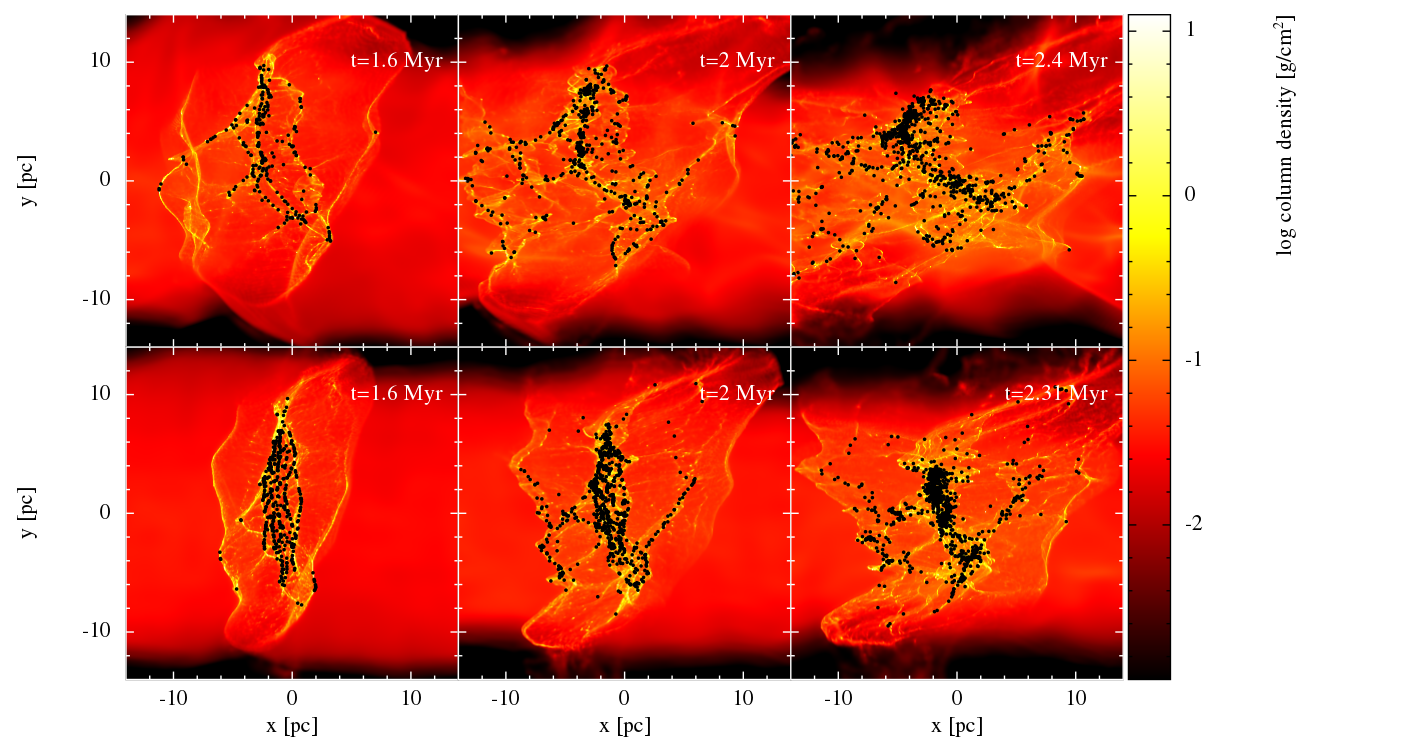}}
\caption{The evolution is shown for the collision of the clouds with the higher virial parameter (BXWHighturb, top panels), and lower virial parameter (BXWLowturb, lower panels). In both cases, the magnetic field is $2.5\times10^{-7}$ G and parallel to the direction of the collision.} 
\label{fig:evolution}
\end{figure*}

In this section we look at the overall evolution for the collisions of clouds with different magnetic field strengths and orientations. We first discuss the results from the simulations with the clouds with higher virial ratios. The evolution of the BWXHighturb model is shown in Figure~\ref{fig:evolution}, top row. The clouds start colliding at around 0.5 Myr. The collision leads to a few main central filaments which are perpendicular to the direction of the collision, as seen in the left panel of Figure~\ref{fig:evolution}.  The shapes, number and structure of these initial filaments are due to the initial turbulent velocity fields of the colliding clouds. These filaments are gravitationally unstable and as such sink particles form along the filaments. As the collision progresses, a more substantial central structure emerges, the number of sink particles increases, and the distribution of sink particles becomes more clustered rather than filamentary. At the final time frame, 2.4 Myr, the sink particles become particularly concentrated to the uppermost region of the collision interface, leading to a more evident cluster here.

\begin{figure*}
\centerline{\includegraphics[scale=0.6]{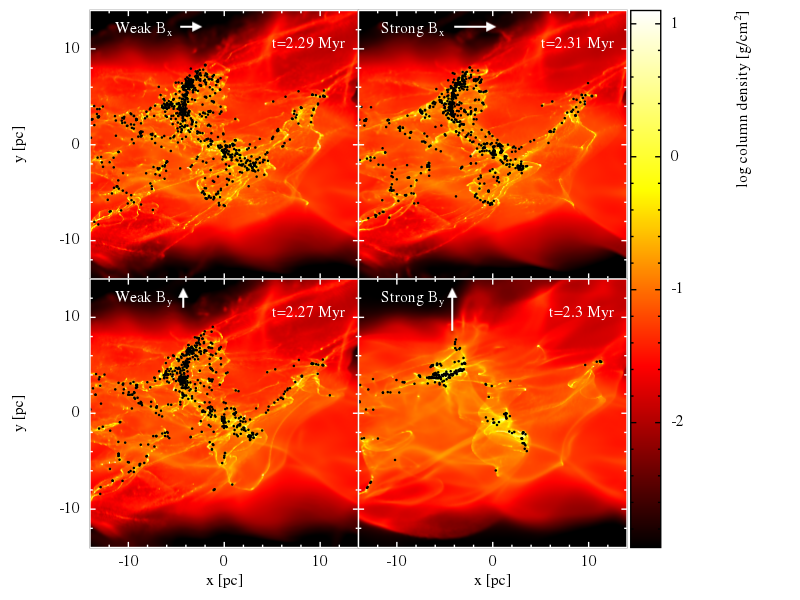}}
\caption{The column density, and distribution of sink particles is shown for the collisions with higher virial parameter clouds. The magnetic field is parallel (top row) and perpendicular (lower) to the direction of the collision, and initial field strength is weaker ($2.5\times10^{-7}$ G) in the left panels, and stronger ($2.5\times10^{-6}$ G) in the right hand panels.} 
\label{fig:highvir}
\end{figure*}

\begin{figure*}
\centerline{\includegraphics[scale=0.6, bb=50 180 800 800]{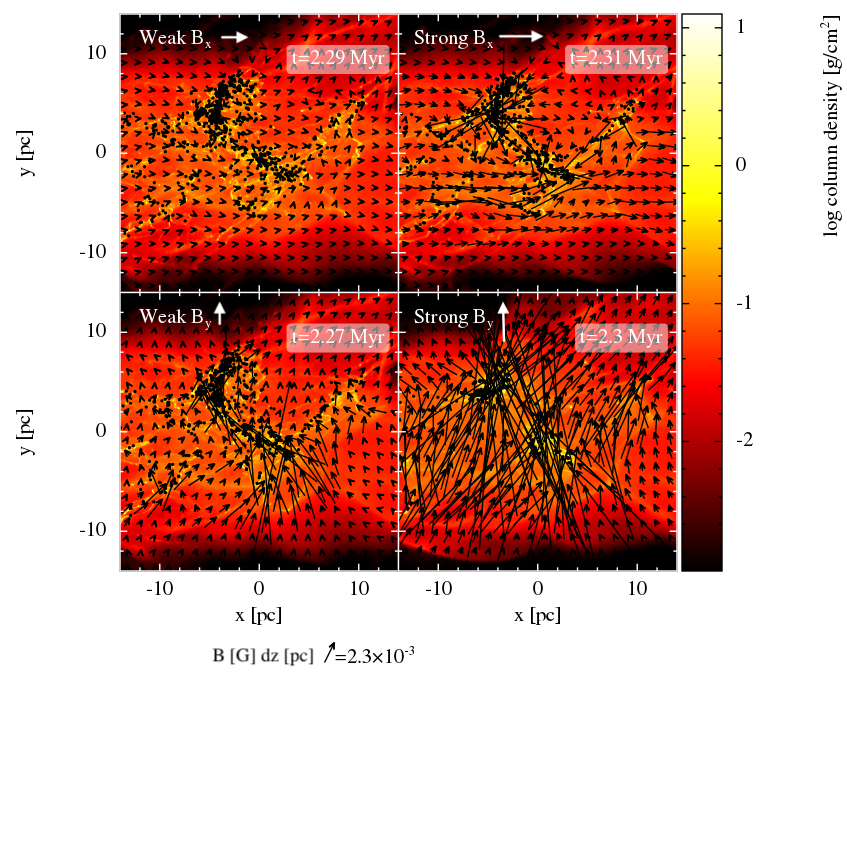}}
\caption{This figure shows the same as Figure~2,but with the magnetic field vectors overlaid.} 
\label{fig:magnetic}
\end{figure*}

\begin{figure*}
\centerline{\includegraphics[scale=0.4, bb=400 100 600 600]{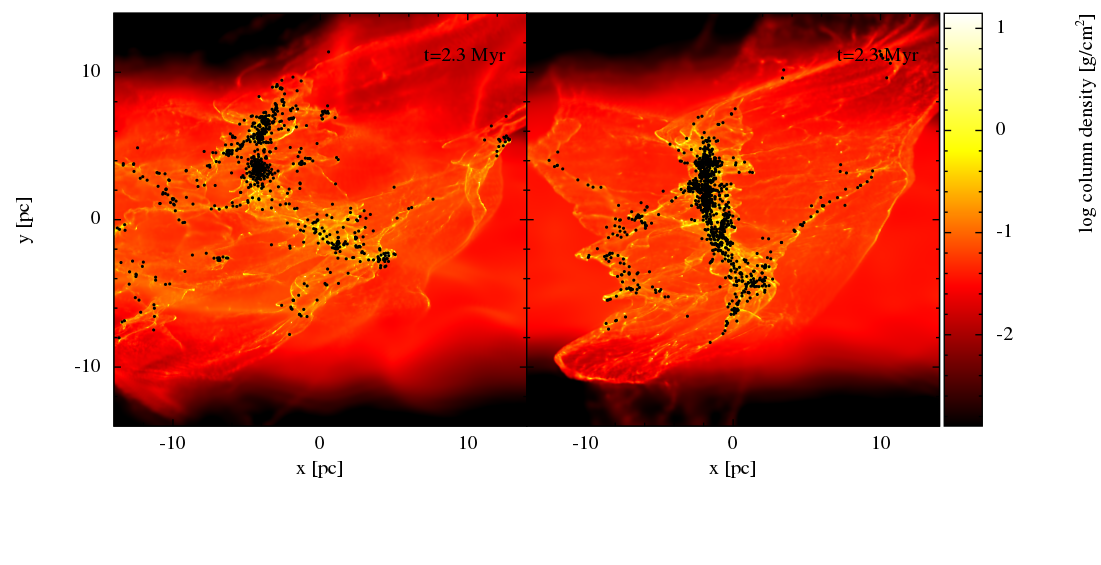}}
\caption{The collisions of clouds in the purely hydrodynamical models is shown for the higher virial parameter clouds (left) and lower virial parameter clouds (right). Both look similar to the MHD simulations, except when the field is stronger and perpendicular to the collision.} 
\label{fig:hydro}
\end{figure*}

The evolution of the other simulations with higher virial parameter is very similar to that shown in Figure~\ref{fig:evolution}, with the exception of the run with a strong field perpendicular to the collision (BSYHighturb). The morphology of the collision for these MHD runs is shown in Figure~\ref{fig:highvir}, at a time of 2.3 Myr. As shown in Figure~\ref{fig:highvir}, the simulations with fields parallel to the collision (aligned along the filament) show very similar morphologies (top row), as does the simulation with a weak field perpendicular to the collision (lower left), although there are some small differences in the morphology of the gas, and the spatial distributions of the sink particles. However the model with a strong field perpendicular to the collision (BSYHighturb) shows a very different morphology. Here the presence of a filament along the collision interface is less clear, and instead sink particles are strongly grouped into distinct clusters. Most of the sink particles are congregated in a cluster in the upper region where the clouds have collided. The evolution of the star formation in this model is also quite different compared to the others, with fewer stars forming earlier compared to the other simulations. 

We show the magnetic field for these models at the same time frame as Figure~\ref{fig:highvir} in Figure~\ref{fig:magnetic}. As expected, the field is stronger in the models where the initial field strength is higher. However we also see that the field has evolved to higher values in the case where the field is originally perpendicular to the collision. The field is clearly strongest, and has been considerably more amplified in the simulation with the strong field perpendicular to the collision. It seems likely that the high magnetic field in the shocked region where the clouds collide is the reason for the difference in morphology, and the resulting difference in star formation. By contrast the model with a weak magnetic field parallel to the collision (BWXHighturb) shows little amplification of the field in the denser, shocked regions. The field also becomes more disordered in the region of the shock. The field becomes most random in the cases where it is initially parallel to the shock, and the shock appears to increase the component of the field along the shock whereas in the models where the field is already aligned with the shock, the effect is more simply to amplify the field in that direction.

We show the equivalent simulation without magnetic fields, HydroHighturb, in Figure~\ref{fig:hydro} (left panel). The morphology of the gas and distribution of sink particles is fairly similar to the models with weak magnetic fields, and the model with a strong field parallel to the shock, although the sink particles appear more concentrated to one main cluster in the hydrodynamical model compared to the MHD cases. The concentration of sink particles into one cluster is more similar to the model with the strong field perpendicular to the shock, BSYHighturb, although otherwise the morphology of the gas and stars is quite different, and there appears to be much more star formation and many more sink particles in the hydrodynamical model (and indeed the other MHD models) compared to the BSYHighturb model.

We now present results for the models where the clouds have lower virial parameters. In Figure \ref{fig:evolution} (lower row) we show the equivalent evolution for the simulation with a low magnetic field parallel to the direction of the collision (BWXLowturb).  Compared to the case with the higher virial parameter clouds, there is a clearer shocked region, and clearer filaments where sink particles are forming. The morphology of the gas, is not too dissimilar to previous grid code simualtions (e.g. \citealt{Kortgen2015}). The distribution of sink particles shows a clear elongated structure. In the last panel, the cluster of sink particles appears to have contracted and is less elongated in the direction perpendicular to the collision, due to gravity acting on the sinks and gas. The evolution of this model appears fairly similar to those presented in \citet{Liow2020} with $\alpha<1$. Similar to \citet{Liow2020}, with a higher virial parameter, the sinks are more dispersed, although in the models here there is still a fairly clear cluster at the location of the shock interface.
\begin{figure*} 
\centerline{\includegraphics[scale=0.6]{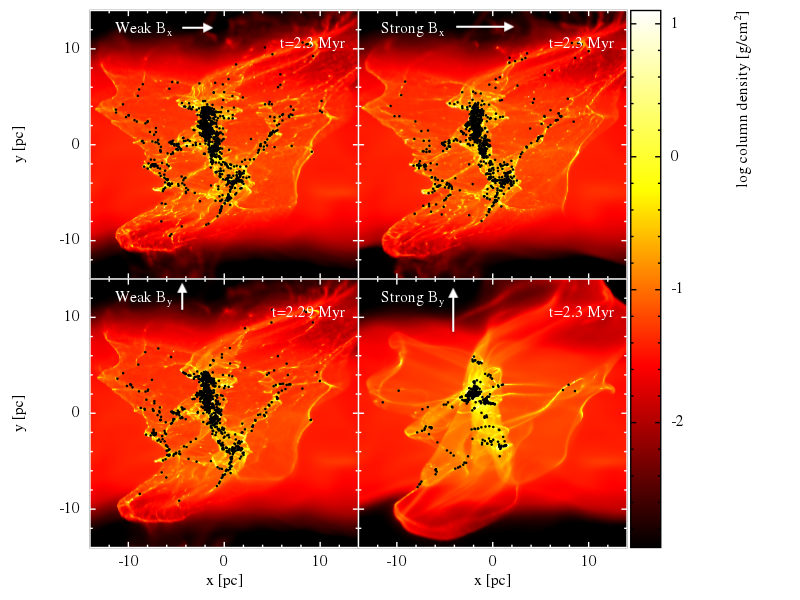}}
\caption{The column density, and distribution of sink particles is shown for the collisions with lower virial parameter clouds. The magnetic field is parallel (top row) and perpendicular (lower) to the direction of the collision, and initial field strength is weaker ($2.5\times10^{-7}$ G) in the left panels, and stronger ($2.5\times10^{-6}$ G) in the right hand panels.} 
\label{fig:lowvir}
\end{figure*}

In Figure~\ref{fig:lowvir} we show the gas surface density plots for the simulations with lower virial parameters and different magnetic field strengths and directions. Again the morphologies of the gas and the sink particles distributions appear similar for three of the simulations, but different for the case with a strong field perpendicular to the direction of the collision (BSYLowturb). Similar to the higher virial parameter simulations, with the exception of BSYLowturb, there is an elongated distribution of sink particles along the shock. By contrast for the BSYLowturb model, far fewer sink particles appear to form, and they tend to be concentrated into a smaller cluster region. The hydrodynamical model, shown in Figure~\ref{fig:hydro} is very similar to the magnetic field models with the exception of BSYLowturb, suggesting that the morphology of these other MHD models is closer to the hydrodynamical case than BSYLowturb.

We do not show the magnetic field vectors for the lower virial parameter models, however the behaviour of the magnetic field is very similar to that shown in Figure~\ref{fig:magnetic}. The field is strongest in the simulation with a strong field perpendicular to the direction of the collision. There is some amplification of the field for the models with a strong field parallel to the collision, and for the weak field perpendicular to the collision, but the field strength appears relatively unchanged with the weak field parallel to the collision.   

In reality, the magnetic field may not be aligned either directly perpendicular or parallel to the collision. We ran a further model where instead the field is aligned at 45 degrees to the direction of the collision. This model shows behaviour that is in between those presented here, i.e. the morphology and distribution of sinks appears in between the cases with a strong field parallel, and perpendicular to the collision, and the magnetic field is amplified to a level in between these two cases.

\subsection{Star formation rates}
\begin{figure} 
\centerline{\includegraphics[scale=0.35, bb=100 50 500 440]{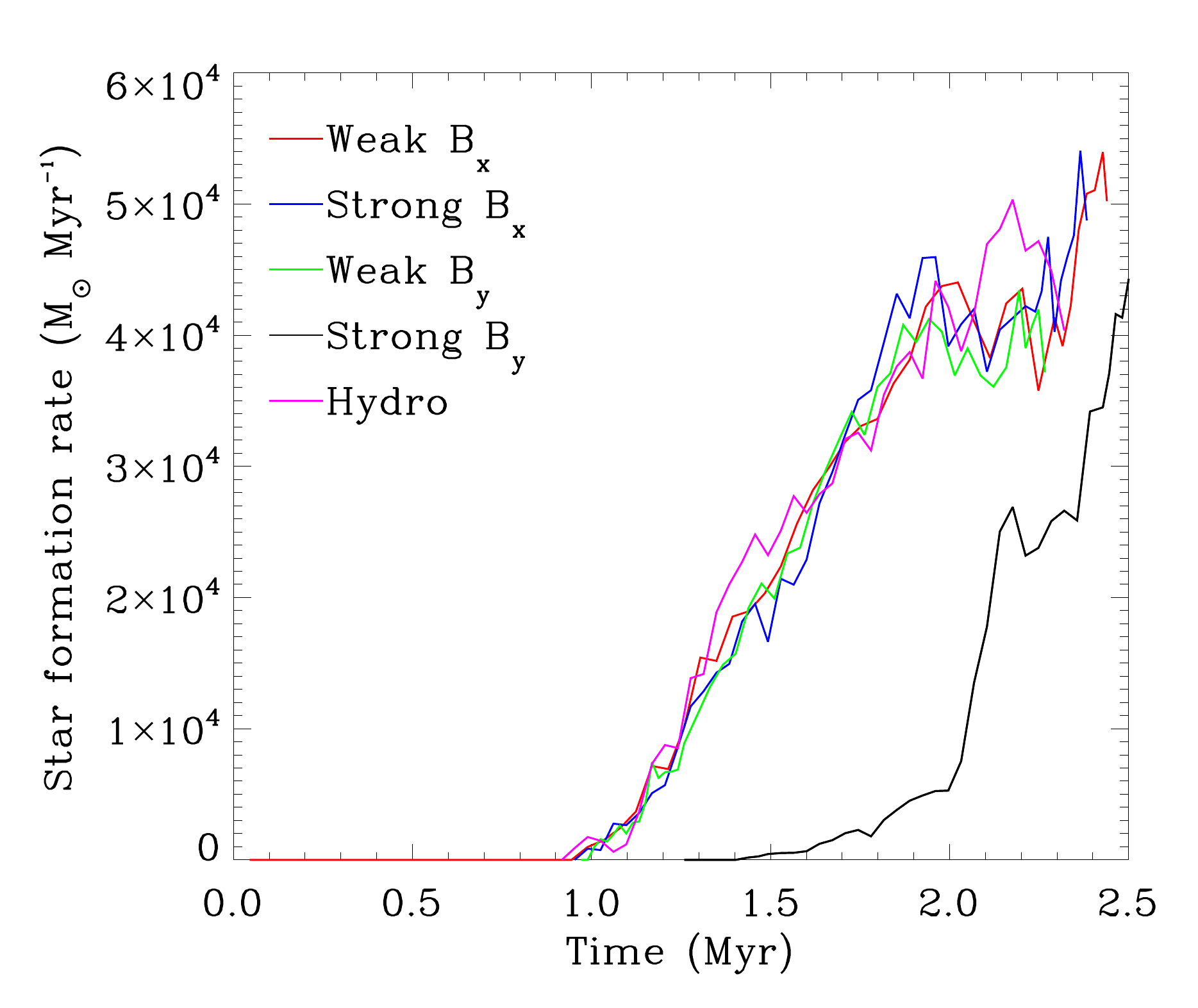}}
\centerline{\includegraphics[scale=0.35, bb=100 0 500 500]{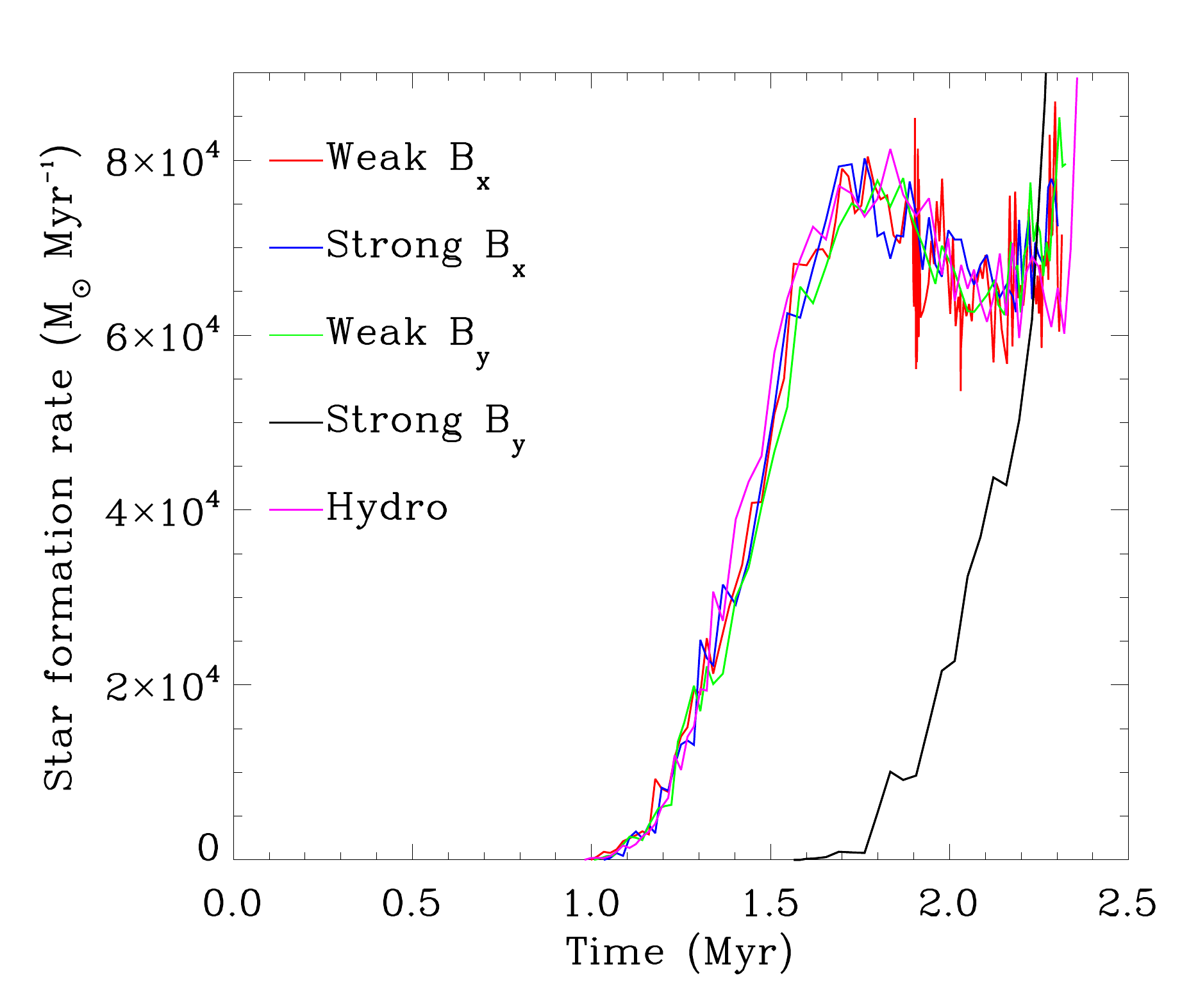}}
\caption{The star formation rates are plotted for the collisions where the clouds have higher virial parameters (top) and lower virial parameters (lower).} 
\label{fig:sfrate}
\end{figure}
We show in Figure~\ref{fig:sfrate} the star formation rates for the different models. The top panel shows the star formation rates in the models with the higher virial parameter. The figure shows that the magnetic field does not appear to impede star formation in most of the models, with the star formation rates extremely similar to the purely hydrodynamical case. This is not so surprising since the morphology of these runs is quite similar. The hydrodynamical model has a slight increase compared to the other models but it is not clear this is particularly significant. The model with a strong field perpendicular to the collision (BSYHighturb) however shows quite a different behaviour in the star formation rate. The star formation increases at $0.5-1$ Myr later compared to the other models. The star formation rate still reaches values as high as the models though, and actually appears to accelerate faster than the other models after the initial delay. We see from Figure~\ref{fig:sfrate} that for the panels in Figures~2-5, shown at a time of 2.3 Myr, star formation has been ongoing for around 1.3 Myr.

In the lower panel of Figure~\ref{fig:sfrate} we show the star formation rates for the models with a lower virial parameter. Again the star formation rates are very similar (almost identical) for all the models except the model with a strong field perpendicular to the collision, even the purely hydrodynamic simulation. This again reflects that the morpohology, and sink distributions of all these models are very similar. Again the model with a strong field perpendicular to the collision (BSYHighturb) shows a delay in the star formation rate, but then the star formation rate increases to values as high, or even higher than the other models. As previously, for a model with a strong field which lies neither parallel or perpendicular to the collision, the star formation rate lies between the BSY models and the other MHD and hydrodynamical models.

\subsection{Magnetic field density relation}
\begin{figure} 
\centerline{\includegraphics[scale=0.35, bb=150 20 500 650]{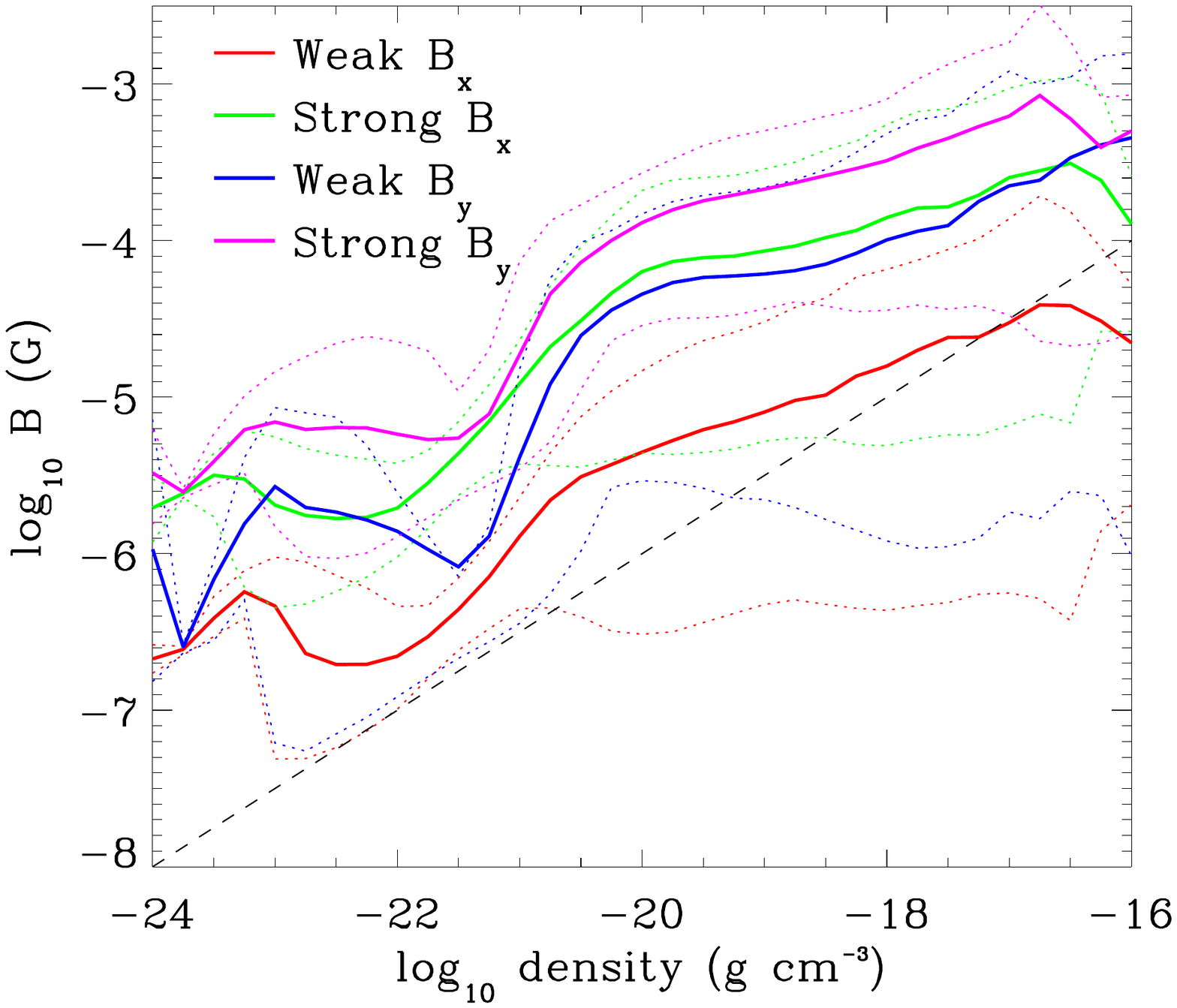}}
\centerline{\includegraphics[scale=0.35, bb=150 170 500 500]{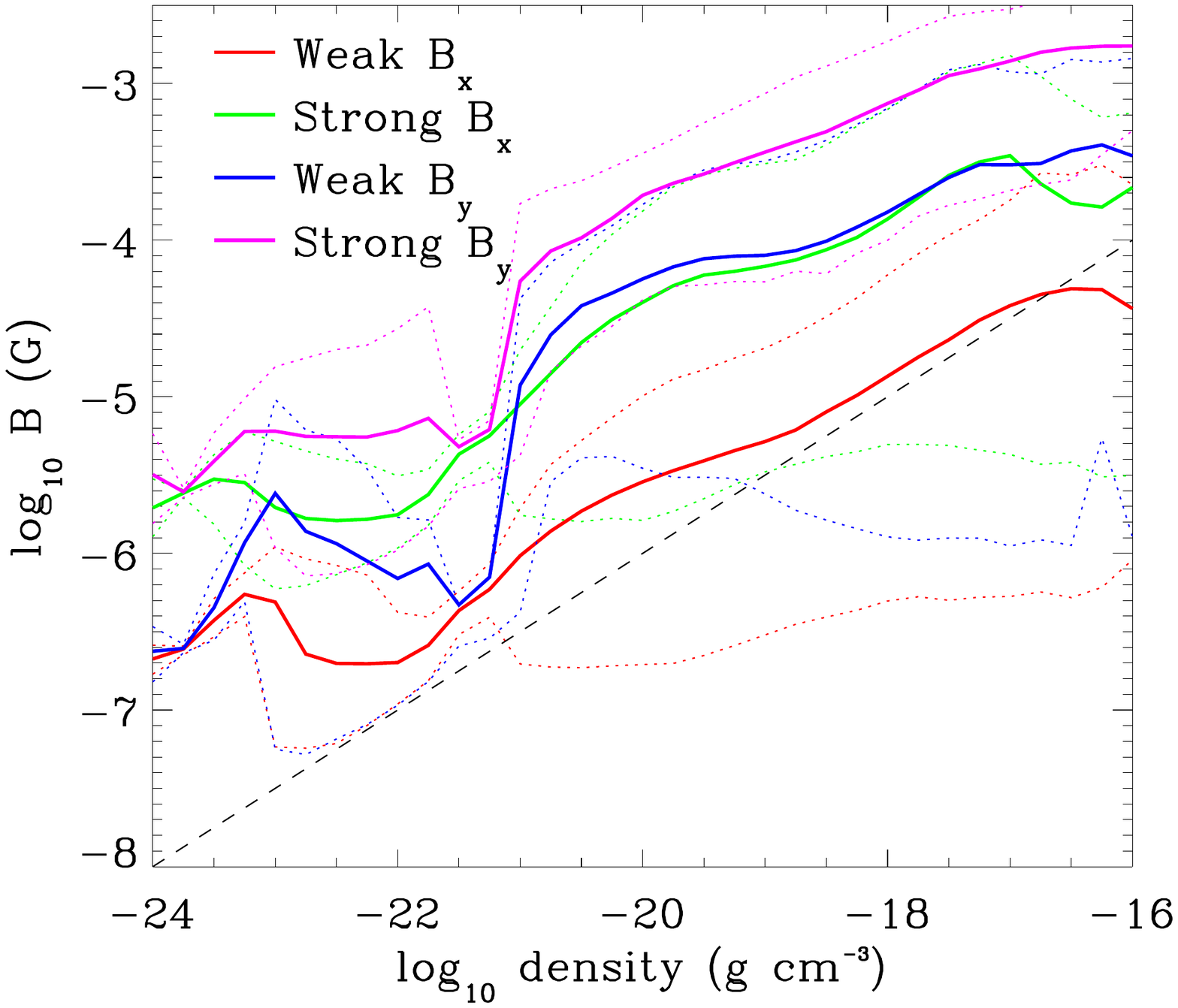}}
\caption{The magnitude of the magnetic field is plotted against density for the collision of clouds with virial parameters (top) and lower virial parameters (lower). The dashed line shows a $B\propto \rho^{1/2}$ relation. The dotted lines show the 95th and 5th percentile lines.} 
\label{fig:brho}
\end{figure}
In Figure~\ref{fig:brho} we show the magnetic density relation for the higher virial parameter (top) and lower virial parameter models (lower) at the same times as shown in Figures \ref{fig:highvir} and \ref{fig:lowvir}. The figures show a region between densities of $10^{-21}$ and $10^{-17}$ g cm$^{-3}$  where the magnetic field scales with the density with a relation slightly shallower than $B \propto \rho^{1/2}$. The initial cloud densities are $\sim2\times10^{-22}$ g cm$^{-3}$, with the background density around 100 times lower. Below densities of $10^{-21}$ g cm$^{-3}$, the magnetic field is roughly constant with density, although the behaviour is more noisy at low densities. Above densities of $10^{-17}$ g cm$^{-3}$, there are relatively few particles to reliably infer a relation. The behaviour of the magnetic field with density is consistent with both \citet{Mocz2017} and \citet{Wurster2019}, even though they model different environments of a turbulent molecular cloud, and disc formation around protostars in a smaller scale region of a molecular cloud. They find a  $B \propto \rho^{1/2}$ correlation at higher densities, whilst the magnetic field is relatively independent at lower densities. The transition however occurs at lower densities in our simulations, where the gas exhibits a lower range of densities, compared to \citet{Wurster2019}. Similar to \citet{Wurster2019} and \citet{Mocz2017}, we see that the magnetic field density correlation is largely independent of our initial conditions, where we are varying initial magnetic field strength, and the initial level of turbulence. We do see a slight tendency for the relation to extend to lower densities in the case with the weakest field parallel to the collision, which perhaps suggests in environments with relatively weaker fields and lower densities the relation extends to lower densities, in agreement with seeing an offset in the flattening of $B$ at low densities, compared with the results of \citet{Wurster2019} and \citet{Mocz2017}.

Unlike \citet{Wurster2019}, where the magnetic field strengths converge above densities of $10^{-18}$ g cm$^{-3}$, we do see offsets in the magnetic field strength for the different models, although the models with a strong field parallel (BSX) and weak field perpendicular (BWY) to the shock are quite similar. The difference is perhaps not surprising, since the evolution is dominated by a strong shock, and the simulations are far from reaching any equilibrium in terms of the gas density distribution, dynamics, and magnetic field. The field strength is highest when the initial magnetic field is strong, and where the field is perpendicular to the collision and experiences amplification. The models which satisfy both, or neither of these properties, represent the outliers for the range of field strengths we use, and across the full range of magnetic field orientations. 

\subsection{Magnetic field and filaments}
\begin{figure} 
\centerline{\includegraphics[scale=0.43, bb=0 20 650 480]{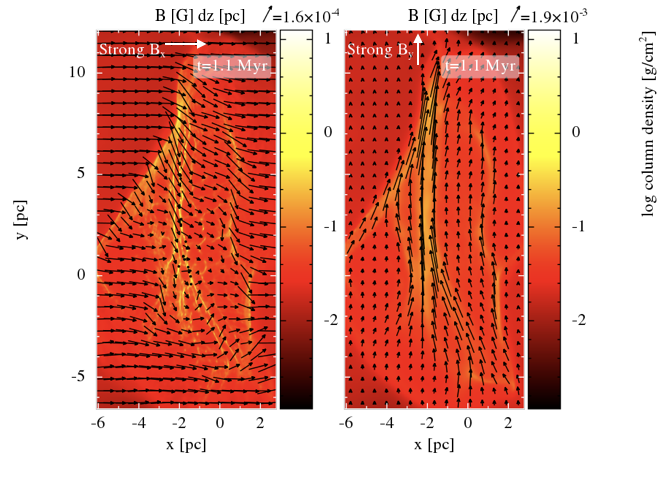}}
\caption{\textbf{These panels show the column density} for the central shocked region (at earlier times in the collision compared to the other figures) for models BSXHighturb (left) and BSYHighturb (right), and the orientaion of the field with respsect to the filaments.} 
\label{fig:filaments}
\end{figure}
In this section we consider further the evolution of the magnetic field, the impact of the magnetic field on the evolution of the collision, and the relation of the magnetic field to the filaments formed at the shock interface. The filaments formed in our simulations typically have lengths in the range 1-10 pc, and large aspect ratios ($>10$). In Figure~\ref{fig:filaments}, we show close ups of the shock region for the models with a strong field initially parallel (left, BSY), and perpendicular (right, BSX) to the direction of collision. Both panels are from the models with the higher virial parameter and higher level of turbulence. At this time (1.1 Myr), the clouds have collided, but there is relatively little fragmentation of the filaments formed from the shock at this point. In the case where the field is perpendicular to the collision, but parallel to the shock, the shock induces an increase in the strength of the field in this direction, as expected for a fast shock \citep{RyuJones1995,Fukui2020b}. Theoretically, we would expect the magnetic field component parallel to the shock to increase by the same amount as the density. We see that both the density and magnetic field strength increase by a factor of several 10's. The large increase in magnetic pressure leads to a broader central filament or shocked region, rather than the narrow dense filaments seen in the other models. Unlike the other models, no sink particles have formed at this point, with the magnetic pressure preventing gravitational collapse (though sink particles do form later).  

In the left hand panel, the field strengths for the model with a strong magnetic field initially parallel to the collision, are much weaker. As expected from theory, there would be no increase in the field if it is perpendicular to the shock (parallel to the collision). As seen by comparing the panels in Figure~\ref{fig:filaments}, the field is more perpendicular to the filament in the left hand panel, but parallel to the filament in the right hand panel. For the left hand panel, there is likely some component of the field parallel to the shock, simply because the turbulent velocity field means the filaments formed from the shock are not completely perpendicular. As such this component will experience some amplification, and in some places the field acquires some component along the direction of the filament. For the left hand panel, the central filament has undergone some fragmentation and a few sink particles have formed along the central filament.

A similar phenomenon whereby the field is parallel to the weaker filaments, and perpendicular to the denser filaments, was noted in Wurster et al. 2019.  Here we look at this a little more quantitatively. We do this simply by selecting the gas in the main central filaments in Figure~\ref{fig:filaments}, and comparing the magnetic field in the $y$ direction with the magnitude of the magnetic field. For the BSX model (left), the density in the filaments increases to $10^{-18}$ g cm$^{-3}$, the $y$ component of the field is on average $\sim 10^{-5}$ G, whilst the magnitude of the field is $\sim 3 \times 10^{-5}$ G. For the BSY model (right), the density in the filaments increases to $10^{-20}$ g cm$^{-3}$, the $y$ component of the field is on average $\sim 1.2 \times 10^{-4}$ G, whilst the magnitude is $\sim 1.5 \times 10^{-4}$ G. 

After the time shown in Figure~\ref{fig:filaments}, the field becomes more aligned with the filaments in the BSX model, but it is not that long before there is more widespread fragmentation, the filaments seen in Figure~\ref{fig:filaments} are less clear, and the field generally becomes more disordered. Similarly for the BSY model, the field becomes more disordered as many sink particles start to form. For the weaker field  models, the field direction is similar to the original orientation of the field. 

In Table~\ref{tab:filament} we have summarised possible outcomes, in terms of magnetic field orientation relative to filaments, and the relative density of the filament, and what initial setup or conditions correspond to this outcome. The Table assumes that the filaments are formed as the result of a high velocity collision, so if the filament formed by an alternative mechanism, it is possible that other inferences could be made. However we see that for our models, a low density filament with magnetic field parallel to the filament occurs if there is a strong magnetic field. The field cannot be perpendicular because that would lead to a high density filament. If the filament is high density, and the field parallel, then the field must be weak, because in this orientation, the field will be amplified, which will lead to a lower density filament if the field is initially strong. For a high density filament with magnetic field perpendicular to the filament, the most likely case is that the field is initially weak, so has little effect on the formation of the filament. Alternatively if the field is strong, the field must be strongly perpendicular to the filament. Interestingly we cannot obtain a relation between magnetic field orientation and filament density, because there is no one-to-one mapping between these two parameters, as in the high density case there are multiple initial conditions which produce a field perpendicular to the filament.  
\begin{table}
\begin{tabular}{ c|c|c| }
\multicolumn{1}{c}{}
 &  \multicolumn{1}{c}{Field parallel}
 & \multicolumn{1}{c}{Field perpendicular} \\
  &  \multicolumn{1}{c}{to filament}
 & \multicolumn{1}{c}{to filament} \\
\cline{2-3}
Low  & Strong field & Not an outcome \\
density & & \\
\cline{2-3}
High  & Weak field & Weak field, or strong field  \\
density & (field unimportant) & with no parallel component\\
\cline{2-3}
\end{tabular}
\caption{Possible outcomes for our simulations are shown according to the column and row labels, and the implication for the field strength.  }\label{tab:filament}
\end{table}

\subsection{Properties of clusters}
\begin{figure*} 
\centerline{\includegraphics[scale=1.0]{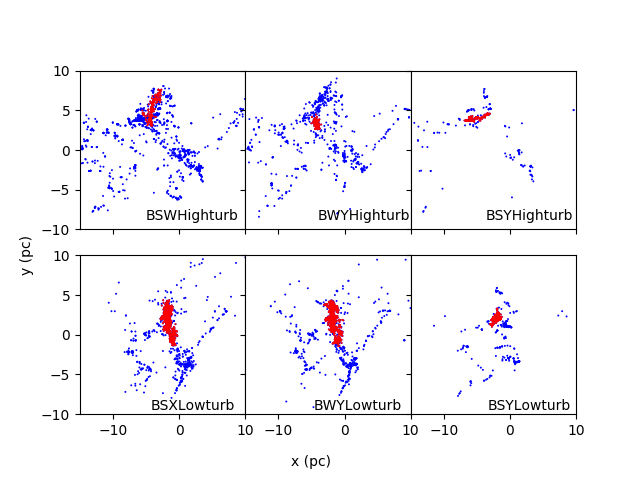}}
\caption{The sink particles are shown for the high virial parameter models (top) and low virial paramters (lower). The red points show the sink particles which are identified as the most massive cluster, as picked out by the DBSCAN algorithm.} 
\label{fig:clustersinks}
\end{figure*}

In this section we study the properties of the clusters formed in the different models. We use the DBSCAN program \citep{Ester1996} to find the 3D distribution of sink particles, as described in \citet{Liow2020}. DBSCAN is a clustering technique which groups together points with similar densities. We use the same maximum separation distance as \citet{Liow2020}, 0.5 pc.
We list the properties of the most massive cluster found in each simulation in Table~\ref{tab:clustertable}. These  properties are listed at a time of 2.3 Myr, which is the same time as shown in Figures~\ref{fig:highvir} and \ref{fig:lowvir}. The properties of the clusters for the clouds with different initial magnetic field configurations are easiest to describe for the lower virial parameter cloud cases (lower 5 entries). We see that the properties of the clusters are very similar for all the models, except that with a strong field perpendicular to the collision (BSYHighturb), where a 5 times smaller cluster has formed. This is not surprising since by eye (Figure~\ref{fig:highvir}), the distribution of sink particles appears very similar in all models except the BSYHighturb model, where the distribution is completely different. By eye, there appear fewer sink particles in the main cluster in BSYHighturb, which although we need to take into account their masses, would suggest a lower mass, smaller radius cluster. In Figure~\ref{fig:clustersinks} we plot the distribution of sink particles for the BSXLowturb, BWYLowturb, and BSYLowturb models, and again it is clear that the BSXLowturb and BWYLowturb models have very similar distributions of sink particles, and likewise the HydroLowturb and BWXLowturb look very similar to these two panels although not shown.

For the models with the higher virial parameter clouds (top 5 rows in Table~\ref{tab:clustertable}), there is more variation in the cluster properties. Again the model with the strong field perpendicular to the collision, BSYHighturb, forms the smallest cluster, which is again unsurprising given the differences in morphology between this and other models, and the lower star formation rates for most of the duration of the simulations. Again, the distribution of sink particles in this model (Figure~\ref{fig:clustersinks}, right panel) is very different from the other models (top left and top centre panels). There is surprising variation in the cluster properties for the other runs. Figure~\ref{fig:clustersinks} shows the sink particle distribution for the BSXHighturb and BWYHighturb models. Here we see that the main accumulation of sink particles towards the top of the panel is identified as a single cluster in the BSXHighturb model, but only a subsection of this region is identified in the BWYHighturb model, hence a less massive cluster is picked out. 

The likely difference between the two sets of models is that for the lower turbulence clouds, the sink particles tend to be located close together for all the models, and the DBSCAN algorithm readily finds a similar mass and size cluster in each case. For the higher turbulence clouds, the higher velocities leads to sink particles, and groups of sink particles which are slightly more disparate to each other. As we see for the BWYHighturb model (Figure~\ref{fig:clustersinks} top middle panel), the large distribution at the top of the panel is separated into about 3 different groups, one of which is picked out as the most massive cluster. In the BSXHighturb (Figure~\ref{fig:clustersinks}, top left panel), and likewise BWXHighturb models, these groups are selected as a single massive cluster. So for the higher turbulence cases, there is a similar distribution of sink particles, but the substructuring within them is slightly different. For the models with the magnetic field parallel to the collision, the substructuring is less pronounced, whereas for the BWYHighturb model with the field perpendicular to the collision, the substructuring is more evident. However it is difficult to say whether the difference is due to the magnetic field, as for the purely hydrodynamical model, the DBSCAN algorithm also only picks out a smaller subcluster (though larger than the BWYHighturb model), or in part due to the random variations in the distributions of the sink particles in different models. 

Overall we see that the addition of a magnetic field tends to make a large difference to the cluster properties if the field is stronger and perpendicular to the collision. Otherwise, if the field is parallel, or weaker, the distribution of sink particles is relatively unchanged on large scales, although we do see differences on smaller scales which may manifest in producing different substructure. For our models with bound clouds (lower virial parameter), we still readily produce massive clusters ($>10^4$ M$_{\odot}$) in all the models, which is probably not surprising given that they are strongly bound clouds. We also see that these clusters form on a timescale of $\sim$ 1.3 Myr. For the higher virial parameter clouds (except for the strong field parallel to the collision, BSYHighturb), we still see large congregations of sink particles with $\sim10^4$ M$_{\odot}$, though they are not always detected as a single cluster by our algorithm. For the strong field parallel to the collision cases, we see less massive clusters, but we note that the star formation is delayed in these instances. If we instead compare at the same time since star formation commences, for the higher turbulence model, the mass of the cluster is 1.4 $\times10^4$ M$_{\odot}$, so more comparable to the other models. However the morphology of the BSY model is still very different, and the resulting most massive cluster is denser and has a smaller half mass radius compared to the other models.

\begin{table}
\begin{tabular}{c|c |c}
 \hline 
Run & Mass (10$^4$ M$_{\odot}$)  & Radius (pc) \\
   \hline
BWXHighturb & 1.62 & 1.2 \\
BSXHighturb & 1.66 & 1.3   \\
BWYHighturb & 0.87 & 0.39   \\
BSYHighturb & 0.33 & 0.12  \\
HydroHighturb & 1.1 & 0.58   \\
\hline
BWXLowturb & 5 & 1.3  \\
BSXLowturb & 5.1 & 1.29   \\
BWYLowturb & 4.96 & 1.35 \\
BSYLowturb & 1.2 & 1.0 \\
HydroLowturb & 5.4 & 1.28 \\
\hline
\end{tabular}
\caption{The properties are listed for the most massive cluster found in each simulation. As for \citet{Liow2020} the radius listed is the half mass radius.} \label{tab:clustertable}
\end{table}

\section{Discussion - comparison with previous work}\label{sec:comparison}
In this section we compare the results of our work to previous simulations. Previous work in this area including magnetic fields has tended to use grid based methods. Our finding of a large dependence on the initial direction of the magnetic field is in agreement with a number of previous works \citep{Heitsch2009,Inoue2009,Fogerty2016}. The earlier works \citep{Heitsch2009,Inoue2009} did not include self gravity and focused on molecular cloud formation, but showed that gas densities only reached values comparable to molecular clouds when the field was parallel to the direction of collision. Similar to our work, the morphology of the gas resembles the hydro case when the field is parallel to the collision, but very different when the field is perpendicular. \citet{Kortgen2015} and \citet{Fogerty2016} also see a delay in star formation with an inclined field compared to a field parallel to the collision, similar to the delay we see.
 Unlike our work, previous simulations see clearer differences with stronger and weaker fields when the field is parallel to the collision \citep{Kortgen2015,Sakre2020,Heitsch2009}, although we do not probe such high magnetic field strengths, where such differences may become more apparent. 
 We also see that in our model with a strong field perpendicular to the collision, fragmentation is strongly compressed, in agreement with \citet{Wu2020}. 

In terms of the mass and time of sinks that form, previous simulations find different results. \citet{Inoue2018} find that the additional magnetic pressure leads to massive stars forming, whereas \citet{Fogerty2016} and \citet{Kortgen2015} find the opposite. \citet{ZA2018} find stars form earlier but \citet{Kortgen2015} and \citet{Fogerty2016} see a delay. We clearly see a delay in agreement with the latter works. Our sink particles represent clusters rather than individual stars. At equivalent times, we see lower mass clusters in the runs with a strong perpendicular field. However if we take the time since the first sinks formed, the situation is less clear, and there is some indication that in the perpendicular case, denser if not necessarily more massive clusters can form. 

Our simulations naturally produce filaments where the shock occurs from the colliding clouds. We see a tendency for magnetic fields to be more aligned with filaments  when the magnetic field impedes the formation of dense gas and stars. The field is instead perpendicular to the filaments formed when the magnetic field has little effect.  The densities of the filaments tend to be lower in the first case, and higher in the second case, in agreement with observations. This finding is comparable to \citet{Soler2013}, who determined the alignment of the magnetic field in different density filaments formed from shocks in turbulent gas, and related the alignment of the field to the divergence of the velocity field \citep{Soler2017}. In our simulations we have a much simpler setup whereby we are modelling individual colliding streams of gas. However we may expect that the outcome for each filament formed in turbulence will have a similar dependence on the initial field strength and angle in our simulations. This idea was approached more rigorously by \citet{Inoue2009} and \citet{Inoue2016} who analytically relate the resultant density of the shock look to the initial angle and strength of the magnetic field prior to a shock. We at least see qualitatively similar behaviour in our models, although we note that there is not necessarily a one to one relationship between filament density and magnetic field properties.

\section{Conclusions}\label{sec:conclusions}
We have performed SPMHD simulations of colliding clouds with magnetic fields to investigate the formation of massive clusters. We apply a timestep criterion which prevents large divergence errors. Although this criteria is not usually necessary in SPMHD calculations, we found it was required in the more extreme conditions of colliding clouds. Our simulations show that magnetic fields do not inhibit the formation of massive clusters from cloud cloud collisions, and YMCs can still form on timescales of 1–2 Myr, and the conclusions from our previous work \citep{Dobbs2020,Liow2020} hold. Even in the case where the impact of the magnetic field is strongest, whilst we see a delay in star formation, we then see star formation rates which are similar to the other models.

Similar to previous work, we find that the initial orientation of the field has a strong effect on the outcome of the collision, and in our case the resultant clusters which form. As shown in \citet{Inoue2009}, this can be related to the conditions of the magnetic field across a shock. If the field is initially parallel to the collision, it has little effect on the evolution even in our stronger field case. Thus dense filaments can form in the gas, and the magnetic field is aligned perpendicular to the filament, as seen in observations. If the field is initially perpendicular to the collision (parallel to the shock), then it is significantly amplified by the shock. As such the magnetic pressure prevents dense filaments forming and delays star formation, and leads to comparably lower density filaments. The magnetic field is aligned along the filament, again in agreement with observations. Thus the formation of the filaments determines the orientation of the field with respect to the filament. Despite the differences with orientation and field strength, we still see a $B \propto \rho^{1/2}$ relation across our models, though the relations are offset from each other.

The influence of the magnetic field on the filaments leads to a corresponding impact on the clusters which form. In the cases where the field has little effect, namely when the field is parallel to the collision, or perpendicular but low strength, the star formation rates and clusters which form have similar properties to the purely hydrodynamic case. In our model where the field has a strong effect (perpendicular, higher strength), the formation of sink particles is delayed, and the resulting appearance and  properties of the clusters which form are quite different. At the same absolute time, the clusters are considerably smaller compared to the other models. At the same duration since star formation commences, the clusters are still less massive though not by so much, but also quite dense. 

Our simulations do not include stellar feedback, which we leave to future work. We would not expect feedback to have a large effect on the gas over the short timescales which our clusters form (e.g. \citealt{Howard2018}), though ionisation may start operating at relatively early times and may also have an impact on the magnetic field \citep{Troland2016}. We also have used fairly modest magnetic field strengths. We would expect the trends we find to continue to higher magnetic field strengths, although we don't necessarily relate our simulations to more extreme environments such as the Galactic Centre, where the dynamics and interstellar radiation field are also very different.

Finally, we are not aware of simulations similar to those presented here which have been carried out with SPH, rather than a grid code. It is encouraging that our simulations produce results, and even filament morphologies, which are in agreement with previous grid code results.

\section*{Data availability}
The data underlying this article will be shared on reasonable request to the corresponding author.

\section*{Acknowledgments}
We thank the referee for providing helpful comments, particularly with regards some of the observations.
We thank Daniel J. Price and Matthew R. Bate for useful discussions regarding the divergence cleaning timestep. We also thank Kong Liow and Alex Pettitt for comments.
Calculations for this paper were performed on the ISCA High Performance Computing Service at the University of Exeter, and used the DiRAC Complexity system, operated by the University of Leicester IT Services, which forms part of the STFC DiRAC HPC Facility (www.dirac.ac.uk ). This equipment is funded by BIS National E-Infrastructure capital grant ST/K000373/1 and  STFC DiRAC Operations grant ST/K0003259/1. DiRAC is part of the National E-Infrastructure. CLD acknowledges funding from the European Research Council for the Horizon 2020 ERC consolidator grant project ICYBOB, grant number 818940.

\bibliographystyle{mn2e}
\bibliography{Dobbs,Wurster}
\bsp
\label{lastpage}
\end{document}